\documentclass[journal=ancac3,manuscript=article]{achemso}
\usepackage[version=3]{mhchem} 

\usepackage{graphicx}
\usepackage{dcolumn}
\usepackage{bm}
\usepackage{braket}
\usepackage{xcolor}
\usepackage{tablefootnote}
\usepackage{array}
\usepackage{setspace}
\usepackage{float}
\usepackage{multirow}
\newcolumntype{C}[1]{>{\centering\arraybackslash}p{#1}}

\newcommand{\red}{\color{red}}
\def\red#1 {\textcolor{red} {#1} }

\title{Environment-Induced Exciton Renormalization in the Photosystem II Reaction Center} 

\author{Tucker Allen}
\email{tuckerallen27@ucla.edu} 
\affiliation{Department of Chemistry and Biochemistry, University of California, Los Angeles, Los Angeles, CA, 90095, USA}
\author{Barry Y. Li}
\affiliation{Department of Chemistry and Biochemistry, University of California, Los Angeles, Los Angeles, CA, 90095, USA}
\author{Nadine C. Bradbury}
\affiliation{Department of Chemistry, Princeton University, Princeton, NJ, 08544, USA}
\author{Daniel Neuhauser} 
\affiliation{Department of Chemistry and Biochemistry, University of California, Los Angeles, Los Angeles, CA, 90095, USA}

\begin{document}

\begin{center}
\textbf{Date: \today}
\end{center}

\begin{tocentry}
\begin{figure}[H]
\centering
\includegraphics[width=3in]{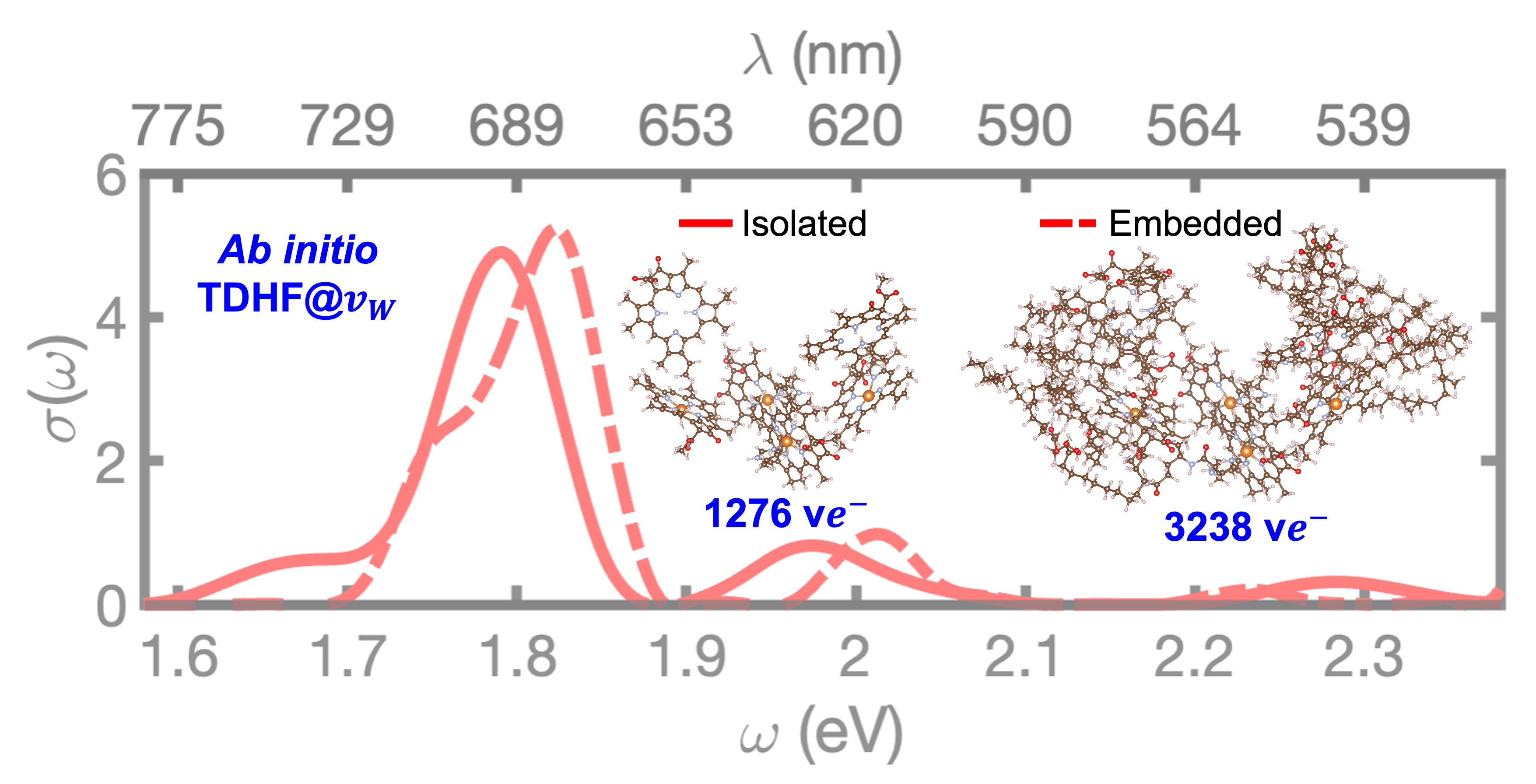} 
\end{figure}
\end{tocentry}

\begin{abstract}
Protein electrostatics tune excitation energies in the Photosystem II reaction center (PSII-RC), yet a fully quantum-mechanical many-body description of how the surrounding protein environment renormalizes excitons has remained computationally inaccessible. The Bethe-Salpeter equation (BSE) within many-body perturbation theory accurately describes excitonic physics through an explicit electron-hole interaction, but is prohibitively expensive for systems containing thousands of valence electrons. Here, we show that for sufficiently large systems the BSE becomes simpler to solve when treated with modern stochastic sampling techniques, as atomistic interactions self-average. In this regime, the effective electron-hole interaction mediated by the environment is governed by collective $k$-dependent polarization. These insights enable an \textit{ab initio} study of the PSII-RC in which all six chlorins forming the hexameric dye core are treated explicitly together with a roughly seven Angstrom local protein environment. We directly compare the low-lying optical excitations of the isolated chromophore hexamer (1276 valence electrons) and the protein-dye cluster (3238 valence electrons). For $Q_y$ excitations near 680 nm, inclusion of the protein environment induces polarization-dependent energy shifts, redistributes spectral weight, and alters exciton delocalization and pigment character. Lateral and transverse asymmetries in the low-lying excited states are captured at the BSE level of theory. These results establish that we now have the tools for many-body calculations of biological nanostructures.

\end{abstract}

Photosystem II (PSII) is a nanoscale molecular machine embedded in the thylakoid membranes of plants, algae, and cyanobacteria. Oxygenic photosynthesis begins at the reaction center of PSII (PSII-RC), where incident photons from sunlight are absorbed by a network of chromophores, generating electron-hole pairs (excitons) and initiating a charge-transfer cascade. Energy transfer drives water splitting and supplies electrons to the photosynthetic electron transport chain. Water oxidation produces molecular oxygen essential for life, as well as reducing equivalents that enable the conversion of carbon dioxide into simple sugars \cite{nelson_complex_2004,stirbet_photosynthesis_2020}. The atomic structure and pigment arrangement of the PSII-RC are now well resolved at near-atomic resolution \cite{Umena2011}. Despite this well-established biochemical pathway and detailed structural knowledge, understanding of the electronic structure of the initial photoexcited states and their sensitivity to the surrounding protein environment remains limited. Further, primary charge separation occurs with near-unity quantum efficiency, making an atomistic, many-body understanding of the PSII-RC essential for uncovering design principles for photovoltaic devices and artificial photosynthesis \cite{nocera_artificial_2012,bredas_photovoltaic_2017}.

Theoretical descriptions of the PSII-RC have historically relied on effective excitonic Hamiltonians that capture energy transfer between pigments in a simplified manner while treating the surrounding protein implicitly \cite{renger_relation_2002,novoderezhkin_physical_2010}. These approaches have provided a useful conceptual picture of exciton energy transfer but neglect the explicit electronic response of the protein environment. More detailed insight has come from mixed quantum-classical approaches, including QM/MM frameworks that employ high-level quantum chemistry methods, such as coupled-cluster theory, alongside range-separated hybrid (RSH) time-dependent density-functional theory (TDDFT) \cite{sirohiwal2020,sirohiwal_electronic_2022,sirohiwal_reaction_2023,dunietz_2025}. These studies have elucidated the role of protein electrostatics, structural disorder, and protein dynamics in shaping excitation energies and charge-transfer pathways. For example, QM/MM studies have shown that the protein electrostatic environment induces transverse and lateral asymmetries that favor charge separation along the D1 branch of the pigment-protein complex \cite{muh_electrostatic_2017,sirohiwal2020}. However, environmental screening is typically described at a classical or semi-classical level and the chromophoric complex is rarely treated as a collective supramolecular electronic system.

For a fully quantum-mechanical treatment, TDDFT has become a widely used tool for excited-state calculations in large photosynthetic complexes. Explicit TDDFT calculations treating the PSII-RC chromophores as a single supramolecular complex were first demonstrated by Frankcombe using the CAM-B3LYP exchange-correlation (XC) functional \cite{frankcombe_explicit_2015}. Subsequent work by Kavanagh \textit{et al.} reported the largest PSII-RC TDDFT model to date, incorporating both the full chromophoric complex and nearby protein residues \cite{kavanagh_tddft_2020}. With highly optimized implementations, TDDFT enables simulations of systems containing thousands of atoms. At the same time, the accuracy of TDDFT and its treatment of many-body effects are fundamentally limited by the choice of XC functional. In heterogeneous systems such as PSII, long-range electron-hole interactions and anisotropic dielectric screening are difficult to capture within local or semilocal approximations to the XC functional, and charge-transfer excitations remain a persistent challenge \cite{Dreuw2004}. RSH XC functionals improve the description of long-range effects but incur substantial additional cost associated with the evaluation of hybrid exchange \cite{YANAI200451,baer_tuned_2010,roi_tddft_2009,Roi_PRL_2010}. Further, the use of a single range-separation parameter across chemically distinct regions of the complex constrains the ability to properly describe screening. In addition, common tuning procedures are not size-extensive, although a recent density-based tuning strategy has been proposed \cite{mandal_simplified_2025}.

Many-body perturbation theory (MBPT), and in particular the Bethe-Salpeter equation (BSE), provides a more explicit treatment of excitonic effects through a screened electron-hole interaction. In simple terms, the BSE, under the static approximation, is equivalent to time-dependent Hartree Fock (TDHF) with a modified exchange term, where the effective interaction \cite{strinati1988,Louie_BSE_2000,Blase2020}: 
\begin{equation}
\label{Woperator}
W(r,r',\omega=0)=\epsilon^{-1}(r,r')v(r-r'),
\end{equation}
replaces the bare Coulomb interaction. However, constructing $W$ via Eq.(\ref{Woperator}) is computationally prohibitive for systems with several thousand valence electrons. Even within the static approximation, building the screened Coulomb matrix requires explicit summations over many occupied and unoccupied states to form the irreducible polarizability, followed by dense matrix operations to obtain the inverse dielectric function. The resulting scaling and memory costs typically restrict BSE calculations to systems far smaller than realistic photosynthetic complexes, although applications to individual chlorophylls and small pigment assemblies have been reported \cite{hashemi_assessment_2021,li_bethe-salpeter_2022}. 

Recent efforts have begun to push the system-size limits of approaches that incorporate screened electron-hole interactions. Work by Förster and Visscher employed a quasiparticle self-consistent GW (qpGW)-BSE approach to model the full PSII-RC with nearly 2000 electrons, using a pair-atomic density-fitting procedure for matrix elements of Eq.(\ref{Woperator}) \cite{forster_quasiparticle_2022}. Further, our recently developed mixed deterministic and sparse-stochastic long-range hybrid TDDFT approach \cite{sereda_sparse-stochastic_2024} enables simulations of similar size and beyond. However, there remains a need to go beyond treating the chromophoric complex alone by incorporating explicit protein environments and simultaneously including electron-hole screening beyond the traditional tuned RSH-TDDFT level.

Here, we adopt an alternative approach that avoids building and storing the screened Coulomb elements explicitly. Specifically, we employ our recently developed screened time-dependent Hartree Fock-like method, TDHF@$v_W$ \cite{no_more_gap}, which replaces the exact screened Coulomb operator, $W(r,r')$, with a translationally invariant screened exchange kernel, $v_W(r-r')$. This assumption reflects the fact that for large systems, it is the overall $k$-dependent polarization, not individual atom-atom interactions, that is important. The effective kernel $v_W$ is constructed via a least-squares fitting procedure, obtained from independent applications of Eq.(\ref{Woperator}) to randomly sampled orbital-pair densities; the expansion converges rapidly with the number of pair densities. Acting by $W$ is efficiently calculated through real-time grid-based stochastic time-dependent Hartree dynamics \cite{Neuhauser2014sGW,vlcek_swift_2018,Allen2024,no_more_gap}. This technique replaces the time-evolution of all occupied orbitals with a small set of stochastic states that are each a random linear-combination of all occupied orbitals. This then provides the $k$-space resolved polarization field that has a non-trivial dependence on $k$, and cannot be captured with conventional RSH functionals.

The translationally invariant $v_W(r-r')$ concept is similar to the bootstrap XC-kernel approach developed by Gross and co-workers for periodic solids \cite{Sharma2011}.. The TDHF@$v_W$ framework has previously yielded accurate optical gaps across a broad class of conjugated molecular systems, including polymethine cyanine dyes as well as planar and curved aromatic hydrocarbons \cite{no_more_gap,param-vw}.

Combined with mixed deterministic and sparse-stochastic compression of exchange integrals \cite{bradbury_deterministicfragmented-stochastic_2023,Allen_prb2025}, TDHF@$v_W$ avoids explicit orbital summations and reduces the cost of screened exchange, enabling fully quantum-mechanical simulations of biomolecular systems with thousands of valence electrons. Further, through an efficient iterative approach to diagonalizing the two-particle Hamiltonian, we go beyond the Tamm-Dancoff approximation and assess the role of resonant-antiresonant coupling in environment-induced exciton renormalization.\cite{RARC_PRB_2009} 

In this article, we make a direct, fully quantum-mechanical comparison between the isolated PSII-RC chromophore hexamer (1276 valence electrons, analogous to the work in Refs. \cite{forster_quasiparticle_2022,sereda_sparse-stochastic_2024}) and the same pigment core embedded in a roughly seven Angstrom environment comprising axial histidine ligands, nearby residues, and protein-bound plastoquinones (3238 valence electrons). This enables us to illustrate clearly how the environment reshapes the low-lying excitons in the PSII-RC. The protein-embedded PSII-RC consists of 1331 atoms and is, to our knowledge, the largest to date model of PSII that is treated with beyond-TDDFT methods. Methodological details are provided in the Supplemental Information (SI). We present in the results section the polarization-resolved optical spectra, exciton energies, transition densities, and participation ratios. These observables quantify how anisotropic environmental screening redistributes spectral weight, shifts excitation energies, and modifies exciton delocalization and pigment character. Finally, in the conclusions section we discuss the implications of the new capabilities to simulate large quantum biological systems, and then discuss future directions.



\section{Results and Discussion}

\begin{figure}[H]
\centering
\includegraphics[width=6.5 in]{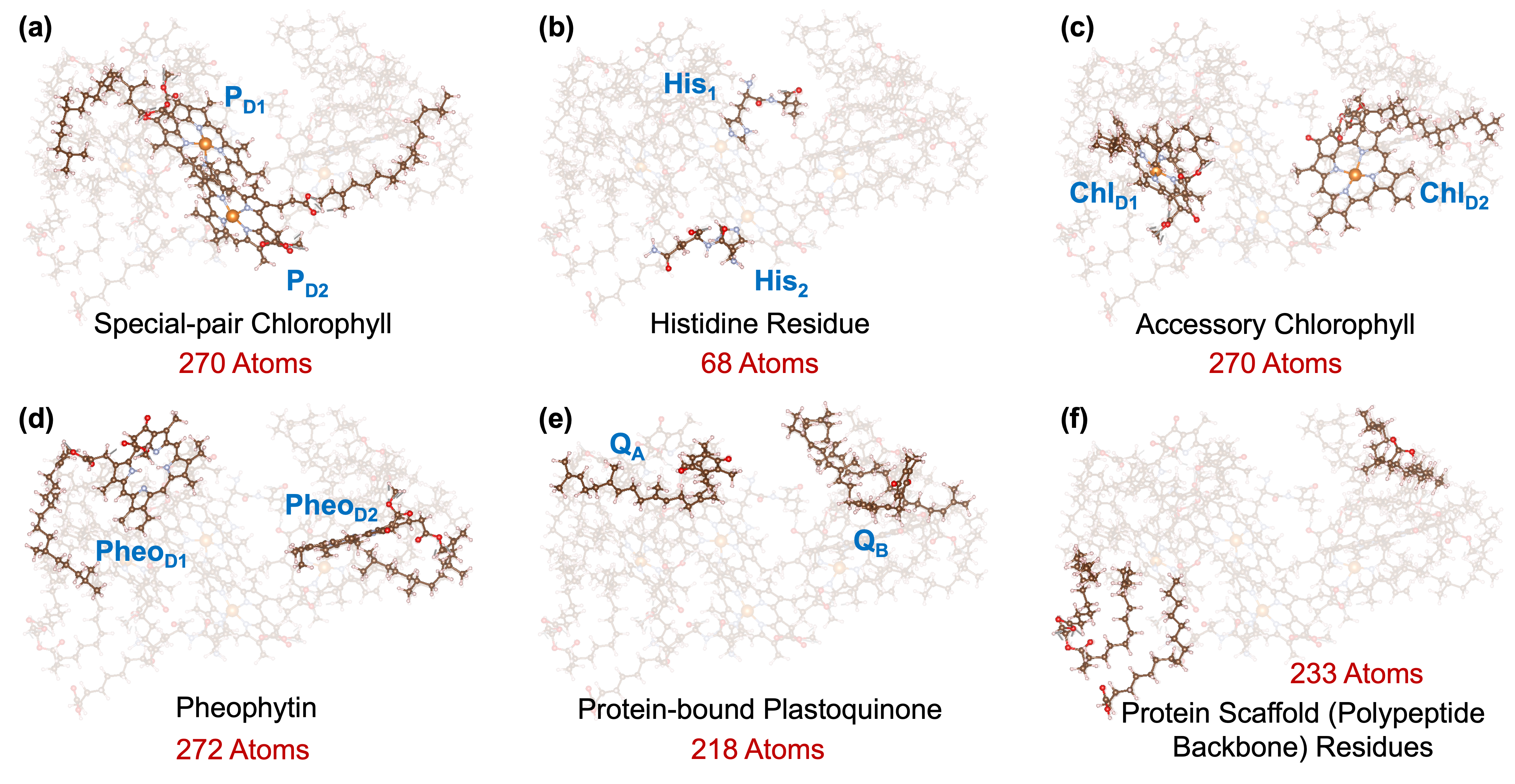}
\caption{\label{fig:segments} Optimized protein-embedded PSII-RC (overall 1331 atoms) with specific segments highlighted: \textbf{(a)} special-pair chlorophylls with Mg centers ($P_{D1}$ and $P_{D2}$), 270 atoms; \textbf{(b)} histidine residues (with backbone) providing axial coordination to the Mg centers of $P_{D1}$ and $P_{D2}$, 68 atoms; \textbf{(c)} accessory chlorophylls with Mg centers ($Chl_{D1}$ and $Chl_{D2}$), 270 atoms; \textbf{(d)} pheophytins with hydrogen-coordinated macrocycles ($Pheo_{D1}$ and $Pheo_{D2}$), 272 atoms; \textbf{(e)} protein-bound plastoquinone cofactors ($Q_A$ and $Q_B$), 218 atoms; \textbf{(f)} surrounding protein scaffold residues (polypeptide backbone), 233 atoms.}
\end{figure}

Fig.~\ref{fig:segments} shows the optimized protein-embedded PSII-RC with explicitly treated pigments, residues, and cofactors, with their names and positions indicated. In Table~\ref{tab:grid_params}, the RSH-DFT bandgaps are provided, and we observe a 0.1 eV gap reduction upon protein embedding.


\begin{table}[H]
\centering
\begin{tabular}{cccccccccc}
\hline
\textbf{System} & $N_x$ & $N_y$ & $N_z$ & $dx$ & $N_{\text{occ}}$ & DFT Gap & $N_v$ & $N_c$ & $N_{k_{\text{low}}}$\\ \hline
Isolated PSII-RC & 136 & 132 & 124 & 0.5 & 638 & 3.51 eV & 200  & 400 & 4683 \\
\hline
Embedded PSII-RC & 168 & 180 & 158 & 0.5 & 1619 & 3.41 eV & 200 & 400 & 10091 \\
\hline
\end{tabular}
\caption{Grid information (an isotropic grid-spacing of $dx=dy=dz=0.5$ Bohr is used), number of occupied orbitals, CAM-LDA0 DFT bandgaps, $N_v$ valence and $N_c$ conduction orbitals used in spectral calculations, and $N_{k_{\text{low}}}$: the number of deterministically treated long-wavelength terms for exchange integrals. The compressed high-$k$ space is represented with 5000 sparse-stochastic vectors, which are sufficient for convergence.}
\label{tab:grid_params}
\end{table}


Stochastic time-dependent Hartree propagation is separately performed for the isolated and protein-embedded PSII-RC. The size of the stochastic basis is $N_\beta=1100$ and $800$, respectively, for the isolated and protein-embedded PSII-RC. Due to self-averaging, the number of samples required to obtain a converged $v_W(k)$ potential tends to decrease with system size.

\begin{figure}[H]
\centering
\includegraphics[width=5in]{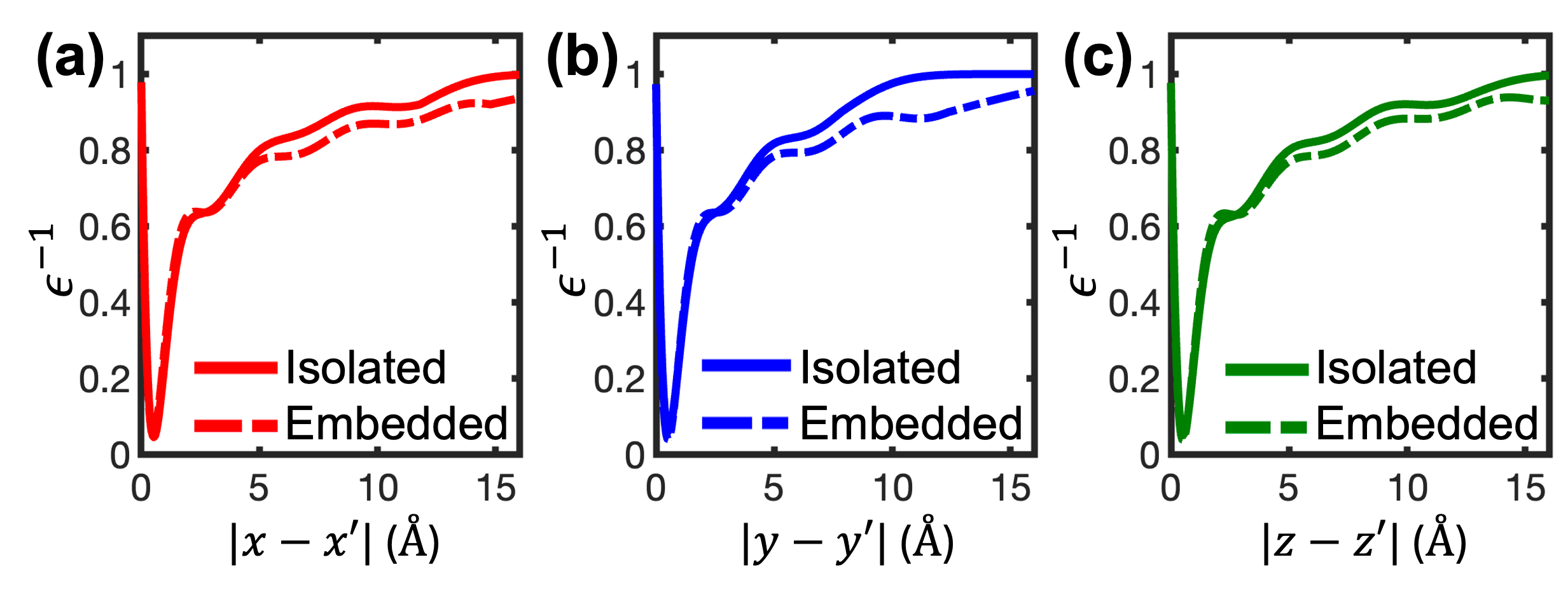}
\caption{\label{fig:vwr} Real-space inverse dielectric functions for the isolated and protein-embedded PSII-RC, evaluated along Cartesian rays from the origin. Panels (a)–(c) show one-dimensional cuts where (a) $|x-x'|$ is varied with $y-y'=z-z'=0$, (b) $|y-y'|$ is varied with $x-x'=z-z'=0$, and (c) $|z-z'|$ is varied with $x-x'=y-y'=0$.}
\end{figure}

From the attenuated exchange kernel, $v_W$, we form the inverse dielectric function, $\epsilon^{-1}(k)=1+v_W(k)/v(k)$, where $v(k)$ is the full (unscreened) non-periodic Coulomb interaction (see SI for details) \cite{MartynaTuckerman1999}. After the Fourier transform, the real-space inverse dielectric is obtained, as shown in Fig.~\ref{fig:vwr}. At short spatial-range, i.e., large reciprocal lattice vector $k$, the screening behavior is very similar between the isolated and embedded systems. At spatial separations of roughly 5 $\text{\AA}$ or more, we observe deviations in exchange attenuation between the two systems. It is at large interelectronic distances (low-$k$) that screening is enhanced in the protein-embedded cluster. 

\begin{figure}[H]
\centering
\includegraphics[width=5in]{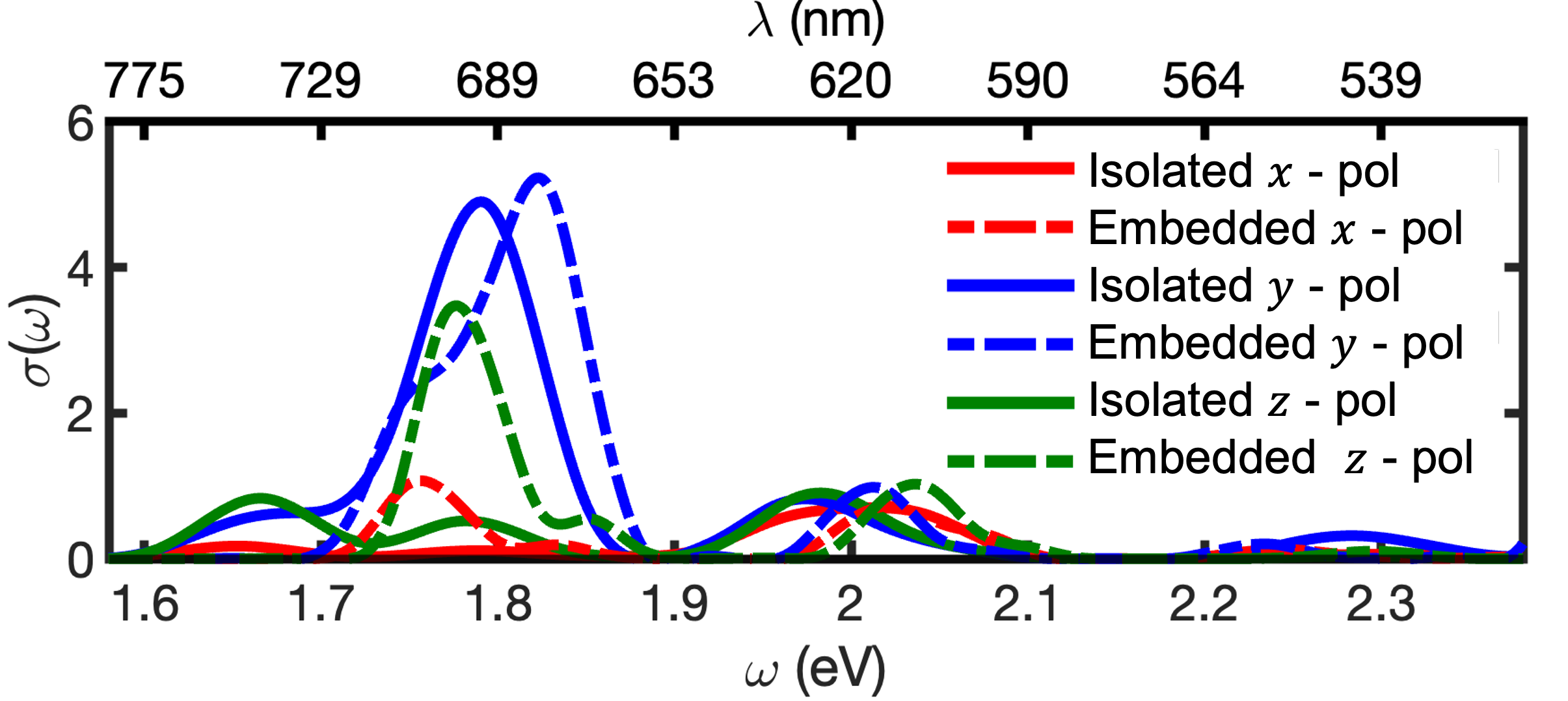}
\caption{\label{fig:spectra} Polarization-resolved full TDHF@$v_W$ optical spectra for the isolated and protein-embedded PSII-RC systems. Both systems use a transition space of  $N_v=200$ and $N_c=400$ orbitals. Laser polarization directions are denoted as $x$-pol, $y$-pol, and $z$-pol.}
\end{figure}

Fig.~\ref{fig:spectra} shows the polarization-resolved optical absorption spectra for the isolated and protein-embedded PSII-RC. To converge the spectral shapes below 3 eV in both systems, it is sufficient to use a transition-space of $N_v=200$ valence and $N_c=400$ conduction orbitals; we have confirmed that going beyond $N_c=400$ does not shift peak positions by more than 0.01 eV. The absolute peak positions are then extrapolated to $N_v=N_{\text{occ}}$ and $N_c=2N_{\text{occ}}$ by a simple linear fitting of the excitation energies as a function of $1/N_v$. Spectra provided in Fig.~\ref{fig:spectra} are shifted to include this extrapolation. Details on the fitting are provided in the SI. 

\begin{figure}[H]
\centering
\includegraphics[width=6.2in]{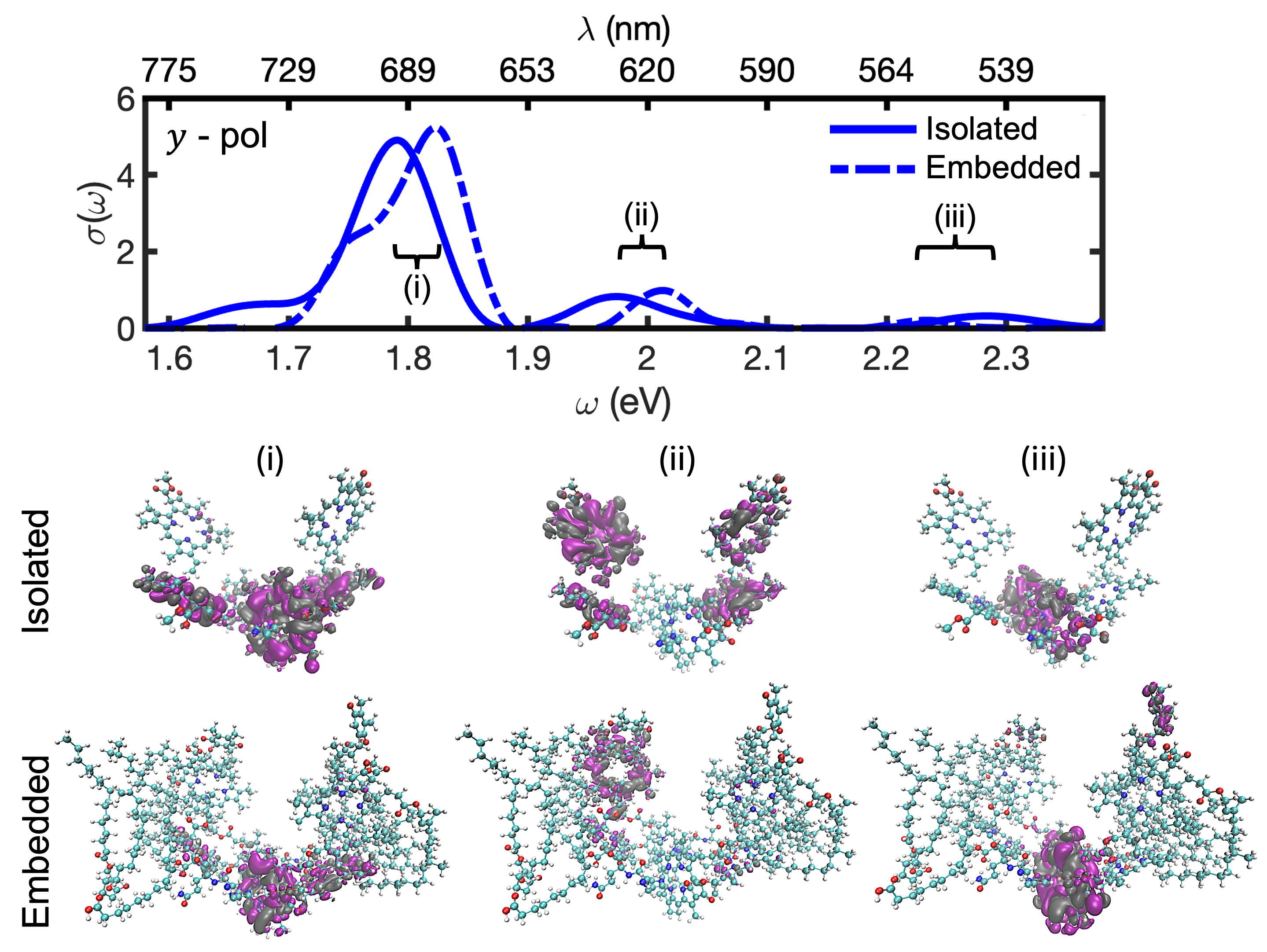}
\caption{\label{fig:densy} The $y$-polarized TDHF@$v_W$ spectra for the isolated and protein-embedded PSII-RC, together with the transition density isosurfaces corresponding to the three dominant absorption peaks in both systems, labeled as \textbf{(i)}, \textbf{(ii)}, and \textbf{(iii)}. Isovalues of $\pm 0.00005$ $\text{Bohr}^{-3/2}$ are used for all transition densities with purple used for positive values and grey for negative values.}
\end{figure}

The three lowest $y$-polarized excited states and associated transition densities are shown in Fig.~\ref{fig:densy}. Table~\ref{tab:exc_chara} summarizes the chromophoric character of each exciton and its participation ratio (PR), which describes the effective number of valence-to-conduction orbital transitions involved for an excitation at a sampled frequency $\omega$. Peak (i) is the bright $Q_y$ band excitation, with a peak position of 1.79 eV for the isolated PSII-RC that is blue-shifted by 0.03 eV upon protein-embedding. We observe a reduction in the PR upon embedding, indicating an environment-induced localization. The theoretical absorption peak of 1.82 eV (681 nm) for the embedded PSII-RC is in excellent agreement with the experimental absorption of 1.83 eV (679 nm) reported in \cite{kavanagh_tddft_2020}. 

\begin{table}[H]
\begin{tabular}{c|cccc}
\hline
\textbf{Exciton} & System & Energy (eV) & Dominant Character & PR \\
\hline
\multirow{2}{*}{\textbf{(i)}}   & Isolated & 1.79 & $P_{D2}/Chl_{D1}/Chl_{D2}$ & 15.3 \\
\cline{2-5}
                                & Embedded & 1.82 & $P_{D2}/Chl_{D1}/Chl_{D2}$ & 8.6 \\
\hline
\multirow{2}{*}{\textbf{(ii)}}  & Isolated & 1.98 & $Pheo_{D1}/Pheo_{D2}/Chl_{D1}/Chl_{D2}$ & 4.4 \\
\cline{2-5}
                                & Embedded & 2.01 & $Pheo_{D1}$ & 3.5 \\
\hline
\multirow{2}{*}{\textbf{(iii)}} & Isolated & 2.29 & $P_{D1}/P_{D2}$ & 3.0 \\
\cline{2-5}
                                & Embedded & 2.23 & $P_{D1}/P_{D2}$, partial $Q_A/Q_B$ & 3.7 \\
\hline
\multirow{2}{*}{\textbf{(iv)}}  & Isolated & 1.67 & $Pheo_{D1}/Pheo_{D2}$ & 5.8 \\
\cline{2-5}
                                & Embedded & 1.77 & $Pheo_{D1}/Pheo_{D2}/P_{D1}$ & 7.9 \\
\hline
\end{tabular}
\caption{\label{tab:exc_chara} Exciton characteristics of the four selected optical excitations (labeled in Figs.~\ref{fig:densy} and \ref{fig:densz}) of the isolated and protein-embedded PSII-RC. Listed are the excitation energies, dominant pigment contributions, and participation ratio (PR).}
\end{table}

Peaks (ii) and (iii) in Fig.~\ref{fig:densy} have lower absorption cross-sections than (i), but exhibit interesting changes in chromophore character upon the inclusion of local protein environment. Peak (ii) shows a delocalized excited state with transition-density amplitude across both pheophytins and accessory chlorophyll pigments. For the embedded PSII-RC, the peak is blue-shifted and the transition-density is more localized on the D1 branch of the complex. This is consistent with protein electrostatics inducing lateral asymmetry that favors the D1 branch of the pigment-protein complex \cite{sirohiwal2020}. Peak (iii) for the isolated PSII-RC has transition-density on the central $P_{D1}/P_{D2}$ chlorophyll pair. With explicit environment, the peak is slightly red-shifted and a non-zero amplitude appears on the $Q_A/Q_B$ plastoquinones. 

\begin{figure}[H]
\centering
\includegraphics[width=3in]{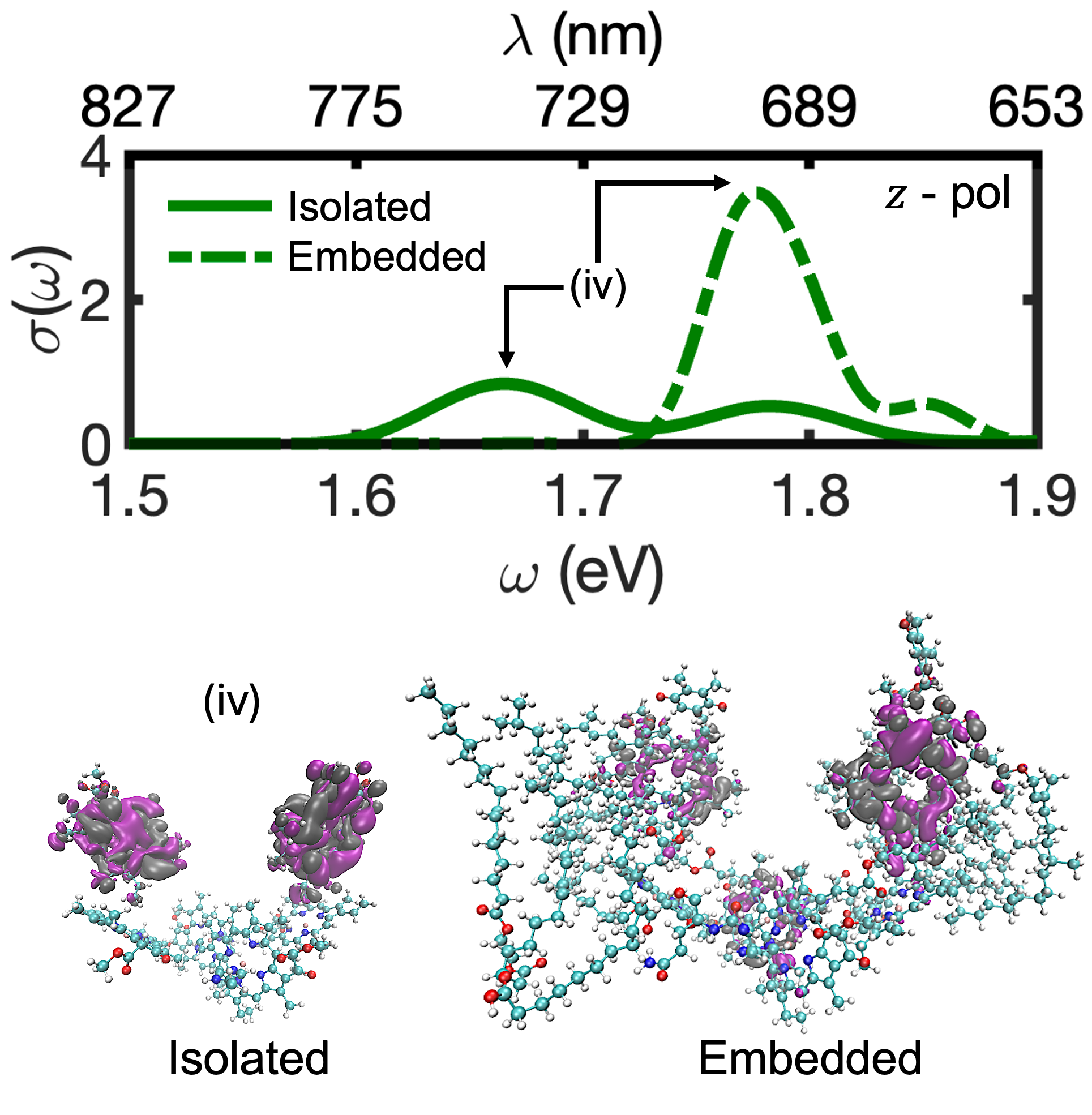}
\caption{\label{fig:densz} Same as Fig.~\ref{fig:densy}, but for the $z$-polarization and the first absorption peak, labeled as \textbf{(iv)}.}
\end{figure}

In Fig.~\ref{fig:densz}, the $z$-polarized absorption onset is provided. A weak doublet is observed in the spectrum for the isolated PSII-RC with transition-density localized on the pheophytins. For the embedded PSII-RC, we observe a 0.1 eV blue-shift and an enhancement in the absorption cross-section of the left peak.  There is a borrowing of spectral weight from higher excitations, with the environment making the transition more dipole-allowed. The PR is slightly increased, and transition-density amplitude appears on the $P_{D1}$ central chlorophyll. 

To assess the role of resonant-antiresonant coupling on exciton renormalization in the PSII-RC, separate Tamm-Dancoff approximation (TDA) TDHF@$v_W$ calculations were performed. In the TDA, peaks (i)-(iv) are rigidly blue-shifted compared with the full TDHF@$v_W$ simulations. Additionally, we do not see noticeable changes in the transition-density characters. However, in the TDA, peak (i) does not blue-shift upon protein embedding. Additionally, peak (ii) does not red-shift. The environment-induced shifts are signatures of transverse anisotropy, and completely missed by the TDA. This signifies that transverse asymmetries in the PSII-RC are attributed to off-diagonal screened exchange interactions. Details of the TDA results are in the SI.  

\section{Conclusions}

In summary, we have presented a many-body BSE study of low-lying excitons in PSII-RC with explicit protein environment using the TDHF@$v_W$ method. Treating the reaction center and its surrounding protein environment on an equal quantum-mechanical footing reveals that the local electrostatic environment does not merely shift excitation energies, but fundamentally reshapes the nature of the excited states. At the \textit{ab initio} level, we show that protein embedding modifies excited states in a polarization-dependent way that alters transition-density character and exciton (de)localization. 

We recover the lateral and transverse asymmetries reported in previous studies \cite{kavanagh_tddft_2020,sirohiwal_reaction_2023}, including D1 lateral branching for the second low-lying state. However, rather than viewing these effects purely as site-energy shifts of individual pigments, our results show that the excited states are inherently multi-pigment and not well captured by a localized picture. A fully quantum, multi-chromophore treatment yields delocalized excitons whose character reflects the collective electronic structure of the complex. The stochastic real-space framework imposes no localization at any stage, so this delocalization emerges naturally.

There is ongoing interest in better understanding the nature of the low-lying charge-transfer (CT) exciton in the PSII-RC. In the present TDHF@$v_W$ calculations, we do not observe a distinct low-energy CT exciton; the low-lying states remain predominantly Frenkel-like. Previous qpGW-BSE calculations on the full chromophoric PSII-RC \cite{forster_quasiparticle_2022} without explicit environmental electrostatics also do not yield a low-lying CT exciton. This confirms that protein conformational dynamics plays a key role in stabilizing and accessing far-red CT states, as earlier suggested by QM/MM studies. \cite{sirohiwal_electronic_2022}.

\subsection{Future Directions}

We have shown that for large supramolecular systems the BSE does not become more intractable, but rather simpler: atomistic interactions self-average, screening becomes polarization-dominated, and stochastic techniques make fully quantum many-body treatment of biological nanostructures possible. Given the demonstrated scalability and accuracy of the present TDHF@$v_W$ framework, several future directions follow:

First, the participation ratio analysis of the excited states indicates that even in very large PSII-RC models only a few tens of occupied-to-virtual transitions (in both isolated and embedded clusters) contribute appreciably to a particular excitation $\omega$. The expansion coefficients, cheaply obtained through the iterative Chebyshev expansion, will enable us to form a minimal exciton basis for effective excitonic models, which will be useful when combined with recent diabatization schemes developed for condensed-phase mixed QM/MM approaches that employ MBPT \cite{rodriguez-mayorga_many-body_2024}. 

A second direction would extend the TDHF@$v_W$ framework beyond the static approximation by incorporating time-dependent screening. Using a frequency-dependent screened Coulomb interaction, $W(r,r',\omega)$, or a stochastically fitted translationally invariant form, $v_W(r-r',\omega)$, within a minimal exciton basis, would capture memory effects associated with the buildup of the electron-hole screening cloud. This would clarify how screening and environmental response influence exciton dissipation, redistribution, and the possible emergence of CT excitons in very large biological complexes \cite{RARC_PRB_2009,Rebolini2016,Loos2020dyn}.


Third, future work will use the present BSE framework to assess the effects of mutating nearby amino-acid residues. The PSII-RC exhibits residue-dependent spectral shifts that can be probed by mutating nearby amino-acid residues. This will be analogous to TDDFT mutation studies done earlier with TD-CAM-B3LYP \cite{kavanagh_tddft_2020}. Coupling the method with molecular dynamics will enable a more realistic treatment of environmental fluctuations, extending the framework toward predictive excitonic structure and dynamics in complex nanoscale photosynthetic systems.

\section{Methods}
\subsection{Computational Details}
Ground-state simulations were performed within a plane-wave pseudopotential DFT framework using Troullier–Martins norm-conserving pseudopotentials \cite{TroullierMartins91} and the LDA exchange–correlation functional \cite{PerdewWang1992}. The mean-field electronic structure was then refined using the near-gap RSH-DFT approach \cite{bradbury_deterministicfragmented-stochastic_2023} with the CAM-LDA0 functional \cite{YANAI200451}. For these conjugated supramolecular systems, the resulting orbital energies provide an accurate starting point for excitonic calculations, and no additional GW quasiparticle correction was applied.

Excited states were computed using the TDHF@$v_W$ framework \cite{no_more_gap}, which is formally equivalent to the static BSE but avoids constructing the screened Coulomb matrix explicitly. Instead, the screened interaction is represented by a translationally invariant kernel, $v_W(r-r')$, obtained from a stochastic fitting procedure. The action of the screened Coulomb operator on randomly sampled occupied–occupied pair densities is evaluated through short-time stochastic time-dependent Hartree propagation \cite{Neuhauser2014sGW,vlcek_swift_2018,Bradbury2022,bradbury_deterministicfragmented-stochastic_2023}. This captures the full $k$-dependent polarization response of the system without building or storing the dielectric matrix.

Exchange integrals in both the RSH-DFT stage and the linear-response calculations were evaluated using our mixed deterministic and sparse-stochastic compression of the plane-wave kernels \cite{vlcek_swift_2018,bradbury_deterministicfragmented-stochastic_2023,sereda_sparse-stochastic_2024}. Long-wavelength components were treated deterministically, while the high-$k$ space was represented by a compact sparse-stochastic auxiliary basis. In practice, this reduces millions of reciprocal vectors to on the order of $10^4$ separable terms at the same level of accuracy.

Optical spectra and frequency-resolved exciton amplitudes were obtained using an iterative Chebyshev expansion of the full two-particle Liouvillian \cite{sereda_sparse-stochastic_2024,no_more_gap}, including resonant-antiresonant coupling beyond the Tamm-Dancoff approximation. This avoids explicit diagonalization of the excitonic Hamiltonian and enables fully quantum-mechanical simulations of biomolecular systems containing several thousand valence electrons. The resulting exciton amplitudes are used to construct the transition-densities provided in the Results section. 

Further numerical details and convergence tests are provided in the SI.



\section{Acknowledgments}
We are grateful to Justin R. Caram for useful discussions. This work is supported by the National Science Foundation (NSF) under Grant No. CHE-2245253. Computational resources were provided by the Expanse Cluster at the San Diego Supercomputer Center through allocations PHY250047, PHY240131, and CHE240067, under the Advanced Cyberinfrastructure Coordination Ecosystem: Services \& Support (ACCESS) program.

\section{Supporting Information Available}
The Supporting Information (SI) includes: methodology; workflow for PSII-RC geometry extraction; basis-set extrapolation of optical spectra; dominant transitions for isolated and protein-embedded PSII-RC; frontier molecular orbitals; parameterized results for the inverse dielectric function $\epsilon^{-1}(k)$; full vs. TDA TDHF@$v_w$ spectra. Cartesian coordinate files (in \text{\AA}) for the isolated and protein-embedded PSII-RC are provided as \texttt{psiirc\_isolated.xyz} and \texttt{psiirc\_embedded.xyz}.

\bibliography{main}
\end{document}